\documentclass[prb,preprint,floats,aps,superscriptaddress,showpacs]{revtex4}
\usepackage{amsmath}
\usepackage{graphicx}

\begin{document}
\date{\today}

\title{Fluid-like behavior of a one-dimensional granular gas}
\author{Fabio Cecconi}
\affiliation{Dipartimento di Fisica, Universit\`a La Sapienza and
INFM UdR Roma-1, P.le A.~Moro 2, I-00185 Rome, Italy}
\author{Fabiana Diotallevi}
\affiliation{Dipartimento di Fisica, Universit\`a La Sapienza,
P.le A. Moro 2, I-00185 Rome, Italy} 
\author{Umberto Marini Bettolo Marconi}
\affiliation{Dipartimento di Fisica, Universit\`a di Camerino,
Via Madonna delle Carceri, I-62032 , Camerino, Italy and
INFM, UdR Camerino}
\author{Andrea Puglisi}
\affiliation{Dipartimento di Fisica, Universit\`a La Sapienza,
P.le A. Moro 2, I-00185 Rome, Italy} 
\affiliation{INFM Center for Statistical Mechanics and Complexity, Italy}      

\begin{abstract}
We study the properties of a one-dimensional
(1D) granular gas consisting of $N$ hard
rods on a line of length $L$ (with periodic boundary conditions).  The
particles collide inelastically and are fluidized by a heat bath at
temperature $T_b$ and viscosity $\gamma$. The analysis is supported by
molecular dynamics simulations. The average properties of the system
are first discussed, focusing on the relations between granular
temperature $T_g=m\langle v^2 \rangle$, kinetic pressure and density
$\rho=N/L$. Thereafter, we consider the fluctuations around the average
behavior obtaining a slightly non-Gaussian behavior of the velocity
distributions and a spatially correlated velocity field; the density
field displays clustering: this is reflected in the structure factor
which has a peak in the $k
\sim 0$ region suggesting an analogy between inelastic hard core interactions
and an effective attractive potential. Finally, we study the transport
properties, showing the typical sub-diffusive behavior of 1D
stochastically driven systems, i.e. $\langle
|x(t)-x(0)|^2 \rangle \sim Dt^{1/2}$ where $D$ for the inelastic fluid
is larger than the elastic case. This is directly related to the peak
of the structure factor at small wave-vectors.
\end{abstract}
\maketitle

\section{Introduction}
Scientists and engineers have been studying granular materials for
nearly two centuries for their relevance both in natural processes
(landslides, dunes, Saturn rings) and in industry (handling of
cereals and minerals, fabrication of pharmaceuticals etc.).
\cite{general1,general2,general3,general4}  
The understanding of the ``granular state'' still represents 
an open challenge and one of the most active research topics in non-equilibrium statistical mechanics and fluid dynamics. For instance, 
a way to attack the problem consists in fluidizing the grains by shaking
them so that the system behaves as a non-ideal gas, a
problem relatively easier to study. The difficulty, but also the beauty,
of the dynamics of granular gases, meant as rarefied assemblies of
macroscopic particles, stems from the inelastic nature of their
collisions which leads to a variety of very peculiar phenomena.
Several theoretical methods have been employed to deal with granular gases 
ranging from hydrodynamic equations, kinetic theories to molecular dynamics.
Engineers often prefer the strategy of the continuum description because it 
gives a better grasp of real life phenomena, while natural scientists tend
to opt for a microscopic approach, to better control each step of the 
modelization. The latter, as far as the interaction between particles is
concerned, regards granular systems as peculiar fluids,
and treats them through the same methods which have been successfully
applied to ordinary fluids.~\cite{hansen}
This allows not only, to employ concepts already
developed by physicists and chemists, but also to stress
analogies and substantial differences.

The purpose of the present paper is to establish such a connection 
for a system of stochastically driven inelastic
hard-rods constrained to move on a ring. The elastic version of this system 
has a long tradition~\cite{lord,tonks,takahashi,frenkel,gursey,salsburg,percus} 
and is particularly suitable to test approximations and theories since
many of its equilibrium properties can be derived in a closed analytical 
form. 
Even though, the one dimensional geometry introduces some
peculiarities not shared by real fluids, we shall
show that the model provides many useful information and a very rich
phenomenology which closely recalls the behavior of microscopic 
particles confined in tubules or cylindrical pores with little 
interconnection.
A second reason to investigate such a model is to show how the
inelasticity of interactions influences not only the
average global properties of a system, but also 
its microscopic local structure.

A basic requirement to a theoretical description of a granular
gas is to provide an equation of state linking the relevant control
parameters and possibly to relate it to the microscopic structure of the
system.  This
connection is well known for classical fluids, where
thermodynamic and transport properties are linked to the 
microscopic level via the correlation function formalism.

One dimensional models have been employed by several authors as simple
models of granular 
gases.~\cite{mcnamara,sela,grossman,zhou,zhou2,kadanoff,mackintosh,puglisi,
puglisi2,bennaim,cordero,baldassarri}
The differences between the various models stem chiefly from the
choice of the thermalizing device. In fact, granular gases would come
to rest unless supplying energy compensating the losses due to the inelastic 
collisions. We call, by analogy, ``{\em heat bath}'' 
the external driving mechanism maintaining the system in a
statistically steady state.

For the history, the first 1D models, which were proposed,
had no periodic boundary conditions and the energy was injected
by a vibrating wall (stochastic or not). This kind of external driving
however, was not able to keep the system homogeneous, because only
the first and last particle had a direct interaction with the
wall.~\cite{kadanoff} As an alternative, a uniform 
heating mechanism, namely a Gaussian white noise acting on each
particle, was introduced.~\cite{mackintosh} Later Puglisi et
al.\cite{puglisi}
added a second ingredient, consisting of a friction term that
prevents the kinetic energy from diverging. With such a
modification the system reaches a steady regime and time
averages can be safely computed.
In the present paper we shall focus on this last model  
characterizing its steady state properties.

The layout is the following: in section~II we introduce the model, in
section~III we obtain numerically and by approximate analytical
arguments equations for the average kinetic energy and pressure. 
Section IV is devoted to fluctuations of the system observables around 
their average values. In section~V we study the diffusion properties of the
system. Finally, in section~VI we present the conclusions.

\section{The model}
Inelastic hard sphere models are perhaps the simplest models able to 
capture the two salient features of granular fluids, namely the hard
core repulsion between grains and the dissipation of kinetic energy
due to the inelastic collisions.  Since many of the equilibrium
properties of the 1D elastic hard rods are known in closed analytical
form, such a system represents an excellent reference model even for
the inelastic case.  Let us consider $N$ identical impenetrable rods,
of coordinates $x_i(t)$, mass $m$ and size $\sigma$, constrained to
move along a line of length $L$. Periodic boundary conditions are
assumed. The hard-core character of the repulsive forces among
particles reduces the interactions to single   
binary, instantaneous collision events occurring whenever
two consecutive rods  
reach a distance $d_i(t)= x_i(t) - x_{i-1}(t)$ equal to their
length $\sigma$.
When two inelastic hard-rods collide, their post-collisional 
velocities (primed symbols) are related to pre-collisional velocities 
(unprimed symbols) through the rule:
\begin{equation}
 v_i' = v_i - \frac{1+r}{2}(v_i-v_j)
\label{collision}
\end{equation}
where $r$ indicates the coefficient of restitution.
The interaction of each particle with the heat-bath is represented by the 
combination of a viscous force proportional to the velocity and a
stochastic force. Then each particle follows the so called Kramers
dynamics
\begin{eqnarray}
 \frac{dx_i}{dt} & = & v_i
\label{kramers1} \\
m\frac{dv_i}{dt} & = & -m\gamma v_i + \xi_i(t)
\label{kramers2}
\end{eqnarray}
where $\gamma$ is the viscous friction coefficient, $\xi_i(t)$
is a Gaussian white noise with zero average and 
correlation 
\begin{equation}
\langle \xi_i(t)\xi_j(s) \rangle  = 2 \gamma m
T_b\delta_{ij} \delta(t-s)\;,
\end{equation} 
$T_b$ is the ``heat-bath temperature'' and $\langle \cdot \rangle$ indicates 
the average over a statistical ensemble of noise realizations.

We have developed a numerical simulation code for hard rods interacting
through momentum conserving but energy dissipating collisions. 
In our simulations the motion between two consecutive collisions
is governed by the dynamics~(\ref{kramers1},\ref{kramers2}).
Thus, we determine the instant when the first collision among the $N$
particles occurs and change their velocities and positions according
to the equations of motion.
The effect of the collision is taken into account by updating the velocities 
after each collision according to the rule~(\ref{collision}).

{We tested our code on the elastic case ($r = 1$)
and checked that our simulations faithfully reproduced the well known
properties of the equilibrium hard rod system.}

\section{Average properties} 
We begin by considering the steady state properties of the model.
The aim is to derive relations connecting the microscopic parameters
to the ``thermodynamic'' observables such as temperature and pressure
and eventually to obtain an ``equation of state'' relating these two
quantities. In order to achieve this goal we assume that the system is
homogeneous, so that its density $\rho$ is constant.

\subsection{Kinetic temperature}
Collisions and the Kramers' dynamics entail that
the time derivative of the average kinetic energy per particle is
\begin{equation}
\frac{d}{dt}\frac{1}{2}m \langle v(t)^2 \rangle =
\gamma(T_b - T_g) - w(t),
\label{collis}
\end{equation}
where $T_g = m \sum_{i=1}^{N} v_i^2/N$ is the granular
temperature and $w(t)$ is the average power dissipated by collisions, 
given by $w = \frac{1-r^2}{8} m \langle \delta v^2 \rangle/\tau_c $,
where $\delta v$ is the difference between the pre-collisional velocities 
of the colliding pair.  
The average collision time $\tau_c$ is estimated by assuming a
mean free path $\lambda = (L-\sigma N)/N$, where $L-\sigma N$ is
the free volume. We obtain, in terms of the system density $\rho = N/L$,
\begin{equation}
\tau_c = \frac{\lambda}{v} = 
\frac{1-\sigma \rho}{\rho}\sqrt{\frac{m}{T_g}}.
\label{tauc}
\end{equation}
Thus the average power dissipated per grain reads:
\begin{equation}
w = \frac{1-r^2}{8}\langle \delta v^2 \rangle \frac{\rho \sqrt{T_g m}}
{1-\sigma \rho}.
\end{equation}
In order to estimate $T_g$ we assume that 
$m \langle \delta v^2 \rangle \simeq 4T_g$
since the velocities of the colliding pairs are strongly correlated.
Thus imposing the solution of Eq.~(\ref{collis}) to be stationary we
obtain for $T_g$ the following expression:
\begin{equation}
T_g = \frac{T_b}{1+\frac{1-r^2}{2\gamma}\frac{\rho}{1-\rho\sigma}
\sqrt{\frac{T_g}{m}}}.
\label{Tg}
\end{equation}
\begin{figure}[htbp]
\begin{center}
\includegraphics[width=8cm,keepaspectratio,clip=true]{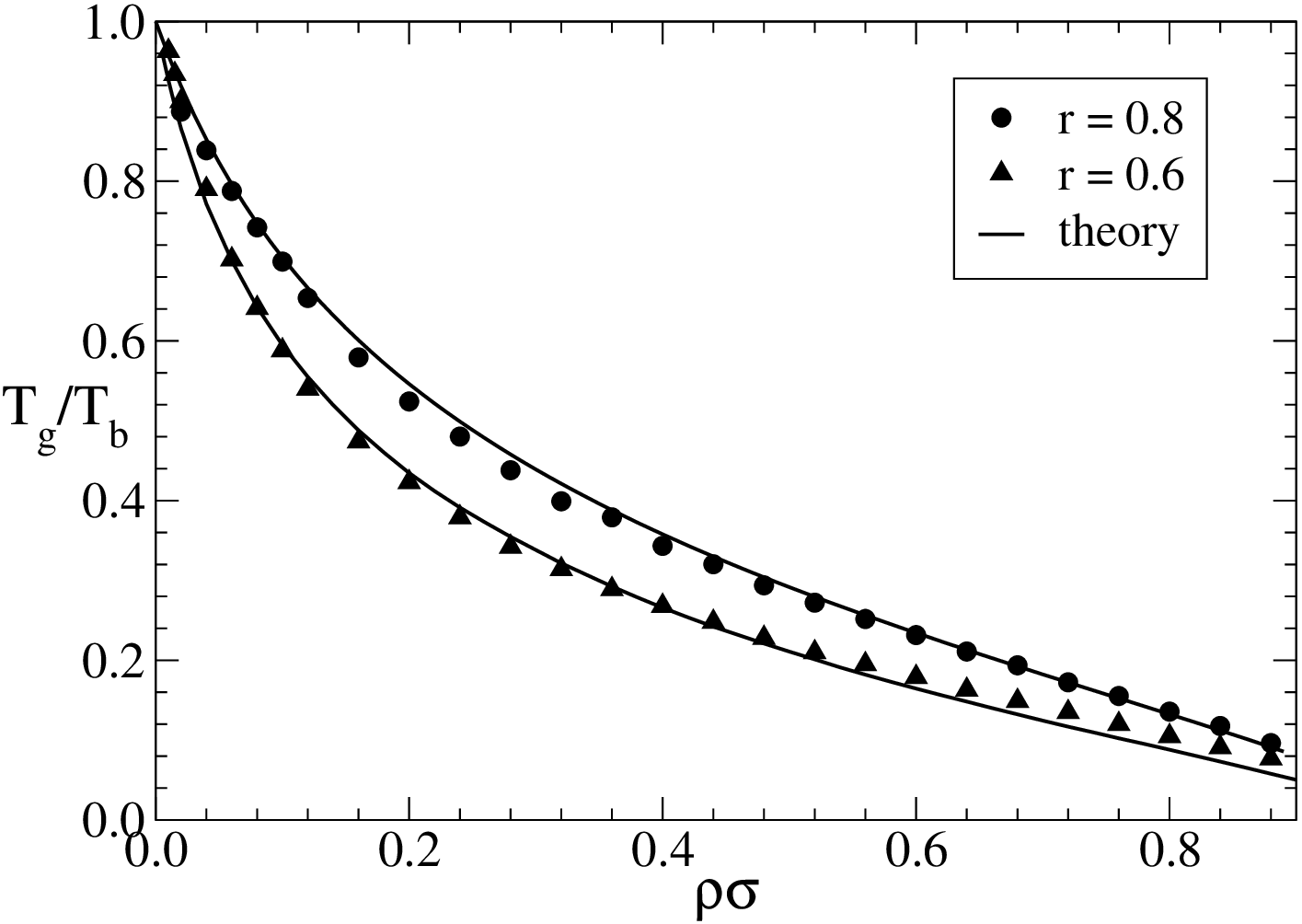}
\end{center}
\caption{Comparison between the numerical results for the
granular temperature versus density and the corresponding
theoretical expression Eq.~(\ref{Tg}). The
simulation data refer to 
$\sigma=0.2$, $\gamma=0.2$, $T_b=1.0$, $r=0.6$ and $r=0.8$. 
We kept the system size fixed to $L=40$ but varied the number of
particles $N$ to change the density.}
\label{fig_graftemp}
\end{figure}
In figure~\ref{fig_graftemp} we compare formula~(\ref{Tg}) with
the results of numerical simulations at various densities.  
In spite of the simplicity of the argument used to derive 
Eq.~(\ref{Tg}), the agreement between
$T_g$ extracted by simulations and its theoretical estimate is 
fairly good.

\subsection{Kinetic pressure}
In a granular system the total pressure, $P$, can be obtained via its
mechanical or kinetic definition, i.e. as the impulse transferred
across a surface in the unit of time.\cite{allen,barrat} 
The pressure contains both the ideal gas
and the collisional contribution ($P_{id}$ and $P_{exc}$ respectively)
\begin{equation} 
P = P_{id} + P_{exc} = \rho T_g(\rho) +
\frac{\sigma}{Lt_{ob}}\sum_{k=1}^{M_c} \delta p_k
\label{Ptot}
\end{equation}
where the second equality stems from the virial
theorem.~\cite{erpenbeck}  Here $t_{ob}$ is the observation time, the
sum runs over the $M_c$ collisions and $\delta p_k = m \delta v_k$
represents the impulse variation due to the $k$-th collision.

An approximate formula for $P_{exc}$ can be derived as follows.    
The average collision frequency per particle can be estimated as
$\tau_c^{-1} = (M_c/t_{ob})/N$.
By replacing in Eq.~(\ref{Ptot}) $t_{ob}$ with $(M_c/N) \tau_c$ and
using $\tau_c$ given by Eq.~(\ref{tauc}), we obtain, 
for the excess part of the pressure, the expression
\begin{equation}
P_{exc} = \frac{N \sigma}{\tau_c L} m \langle \delta v_c \rangle  
= \frac{\rho^2 \sigma}{1-\sigma \rho} T_g(\rho).
\label{excpressure}
\end{equation}

Collecting pieces together we arrive at
\begin{equation}
P(\rho)= T_g(\rho) \bigg[\rho+
\frac{\sigma \rho^2}{1-\sigma \rho}\bigg]\;=\; 
T_g(\rho)\frac{\rho}{1-\sigma \rho}
\label{pressure}
\end{equation}
which reproduces the well known Tonks formula~\cite{tonks} in the
case of elastic particles and constitutes the sought equation of state
for the inelastic system.  Let us recall that
in the elastic case, Eq.~(\ref{pressure}) can be written in the
virial form
\begin{equation}
P(\rho)= T_g(\rho) \rho[1+\rho \sigma g(\sigma)]
\label{pressure2}
\end{equation}
showing the connection between the macroscopic and the
microscopic level, since $g(\sigma)=1/(1-\rho \sigma)$ is the
equilibrium pair correlation at contact.

\begin{figure}[htbp]
\begin{center}
\includegraphics[width=8cm, keepaspectratio, clip=true]{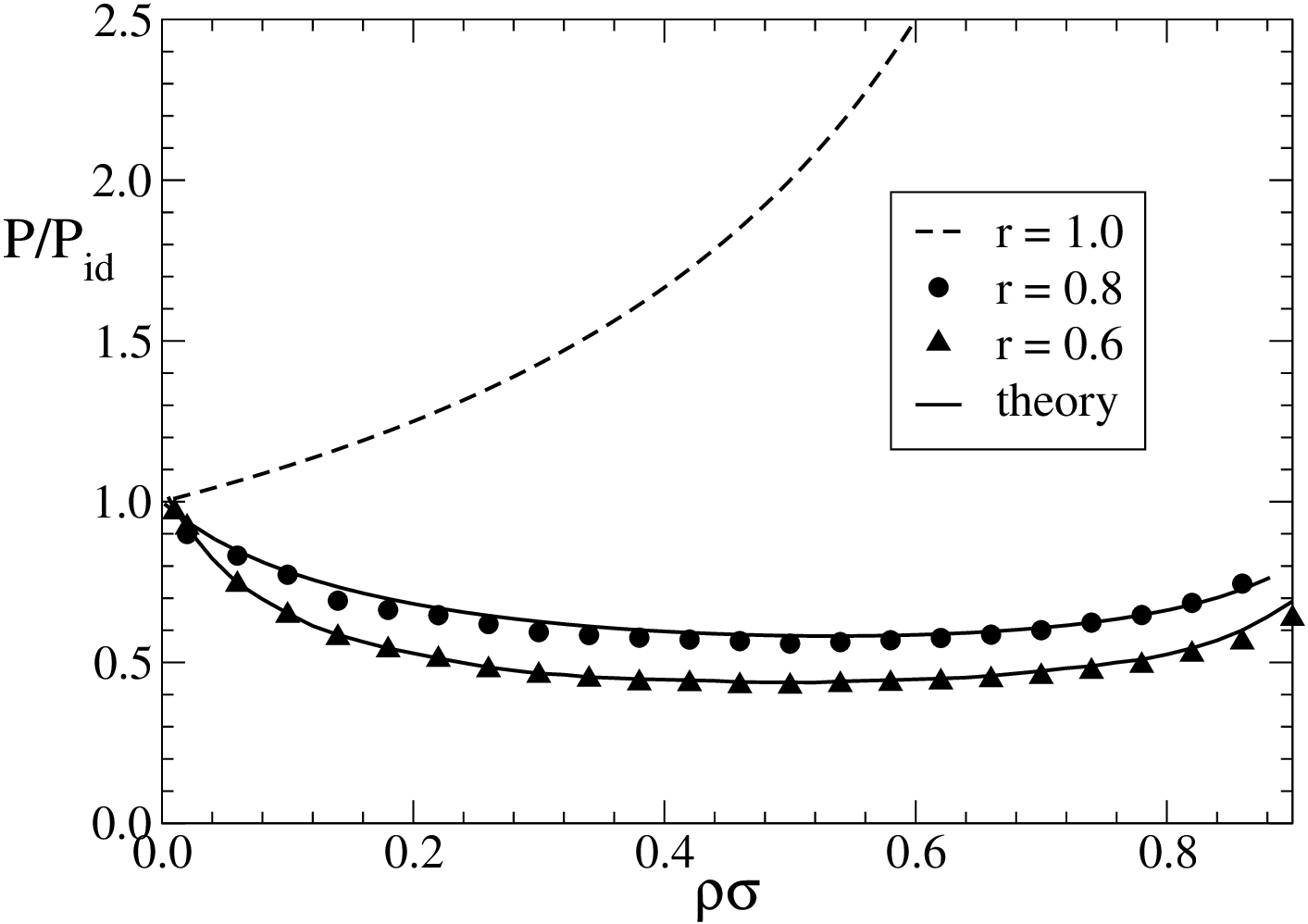}
\end{center}
\caption{Comparison between the
kinetic pressure obtained from the simulations and the
prediction of Eq.~\ref{pressure} at  different densities
for inelastic
hard rods with coefficients of restitution $r=0.6$ and $r=0.8$. The
remaining parameters are the same as in figure~\ref{fig_graftemp}.
Dashed line refers to the pressure of the corresponding 
elastic system.}
\label{fig_pres}
\end{figure}

We see, from figure~(\ref{fig_pres}), that the presence of the prefactor
$T_g(\rho)$, which is decreasing function of the density, makes $P(\rho)$ to 
increase more slowly than the corresponding pressure of the elastic
system in the same physical conditions (i.e. same density and
contact with the same heat bath).  

Equations~(\ref{Tg}) and~(\ref{excpressure}) for temperature and  
pressure coincide, in the limit $\gamma\to 0$, $\sigma\to 0$,  
$\gamma T_b=\Omega=\mbox{const}$, with those derived by    
Williams and MacKintosh~\cite{mackintosh}.  

\begin{figure}[htbp]
\begin{center}
\includegraphics[width=8cm,keepaspectratio, clip=true]{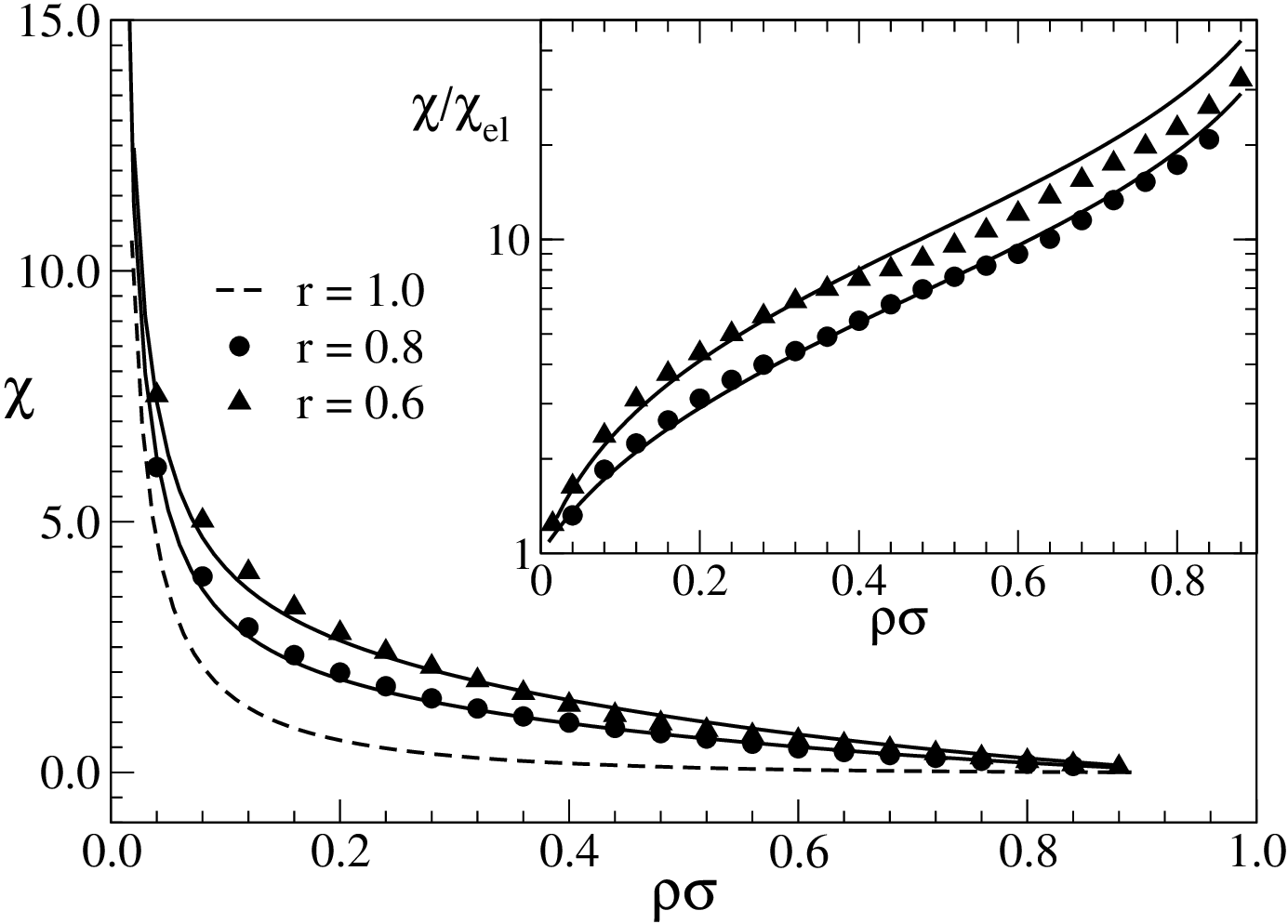}
\end{center}
\caption{Compressibility, computed from
Eq.~(\ref{compressibilita}),  plotted versus density, 
for inelastic hard rods with coefficient $r=0.6$
(the remaining parameters are the same as in
figure~\ref{fig_graftemp}).
Symbols refers to simulations while solid lines are the theoretical
predictions obtained via the expression for the 
pressure~(\ref{pressure}). The curve  for the elastic system 
$r = 1$ (dashed line) is also reported for sake of comparison.
In the inset shows the ratio between the inelastic and elastic
compressibility.}
\label{fig_compres}
\end{figure}

Following the standard approach to fluids, we define, even for the
inelastic system, the response of
the density to a uniform change of the pressure for a fixed value of
the heat bath temperature:
\begin{equation}
\chi_T=\frac{1}{\rho}\frac{\partial\rho}{\partial P}
\label{compressibilita}
\end{equation}
which is plotted in figure~\ref{fig_compres}. We observe that the
response of the inelastic system to a compression is much larger
than the corresponding elastic system at the same density, due to the 
tendency to cluster.

\section{Fluctuations}
So far we have considered only the global average properties of the
granular gas. It is well known, on the other hand, that these system
may exhibit strong spontaneous deviations from their uniform state. In
this section we shall study fluctuations of the main observables in order
to understand the qualitative effect of inelasticity on such a
peculiar fluid.

\subsection{Velocity distributions}
\begin{figure}[htbp]
\begin{center}
\includegraphics[width=8cm,keepaspectratio, clip=true]{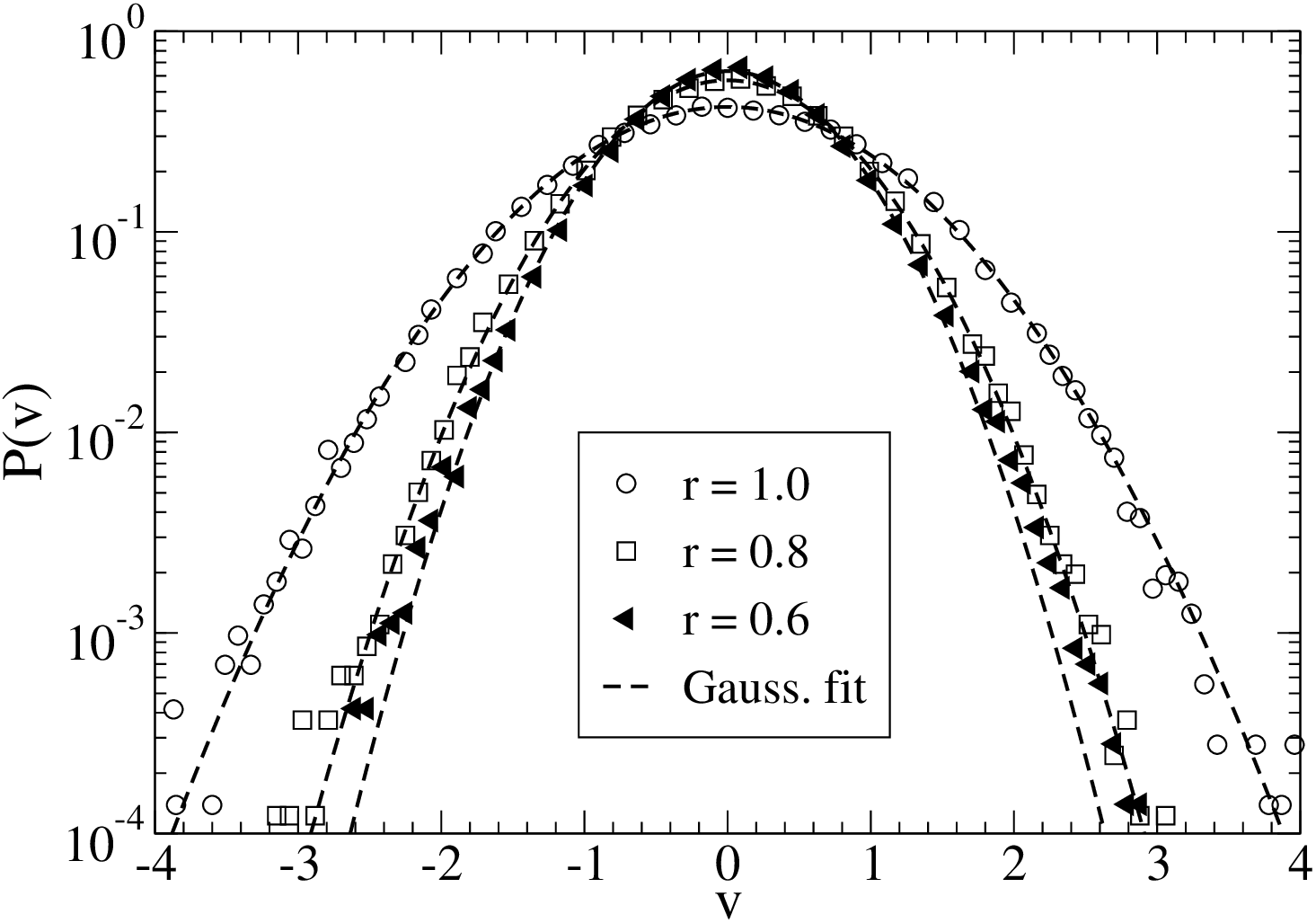}
\end{center}
\caption{
Velocity distributions $P(v)$ for three values of the
coefficient of restitution
$r=1.0$ (circles), $0.8$ (squares), $0.6$ (triangles). 
The remaining parameters are
$N=1000$, $L=1000$, $\sigma=0.2$, $T_b=1.0$, $\gamma=0.2$. Dashed lines 
indicates the corresponding Gaussian fit.}
\label{velocitydistr}
\end{figure}

One of the signatures of the inelasticity
of the collisions is represented by the shape of the velocity
distribution function (VDF), $P(v)$.  
Non-Gaussian VDF's, displaying
low velocity and high velocity overpopulated regions, 
have been measured experimentally \cite{losert,menon,Blair,urbach,aranson,
kudrolli}
and in numerical simulations.\cite{puglisi,breyvel}
In
figure~\ref{velocitydistr} we show two VDF corresponding to two
different values of $r$.

Theoretical, numerical and experimental studies have shown that the
VDF for inelastic ($r<1$) gases usually displays overpopulated
tails. The literature seems to indicate the lack of a universal VDF:
in $d>1$ the solution of the homogeneous Boltzmann equation with
inelastic collisions (with a stochastic driving similar to ours 	but
without viscosity) has overpopulated tails of the kind $\sim
\exp(-v^{3/2})$.\cite{ernst} 

\subsection{Energy fluctuations}
\begin{figure}[htbp]
\begin{center}
\includegraphics[width=8cm,keepaspectratio, clip=true]{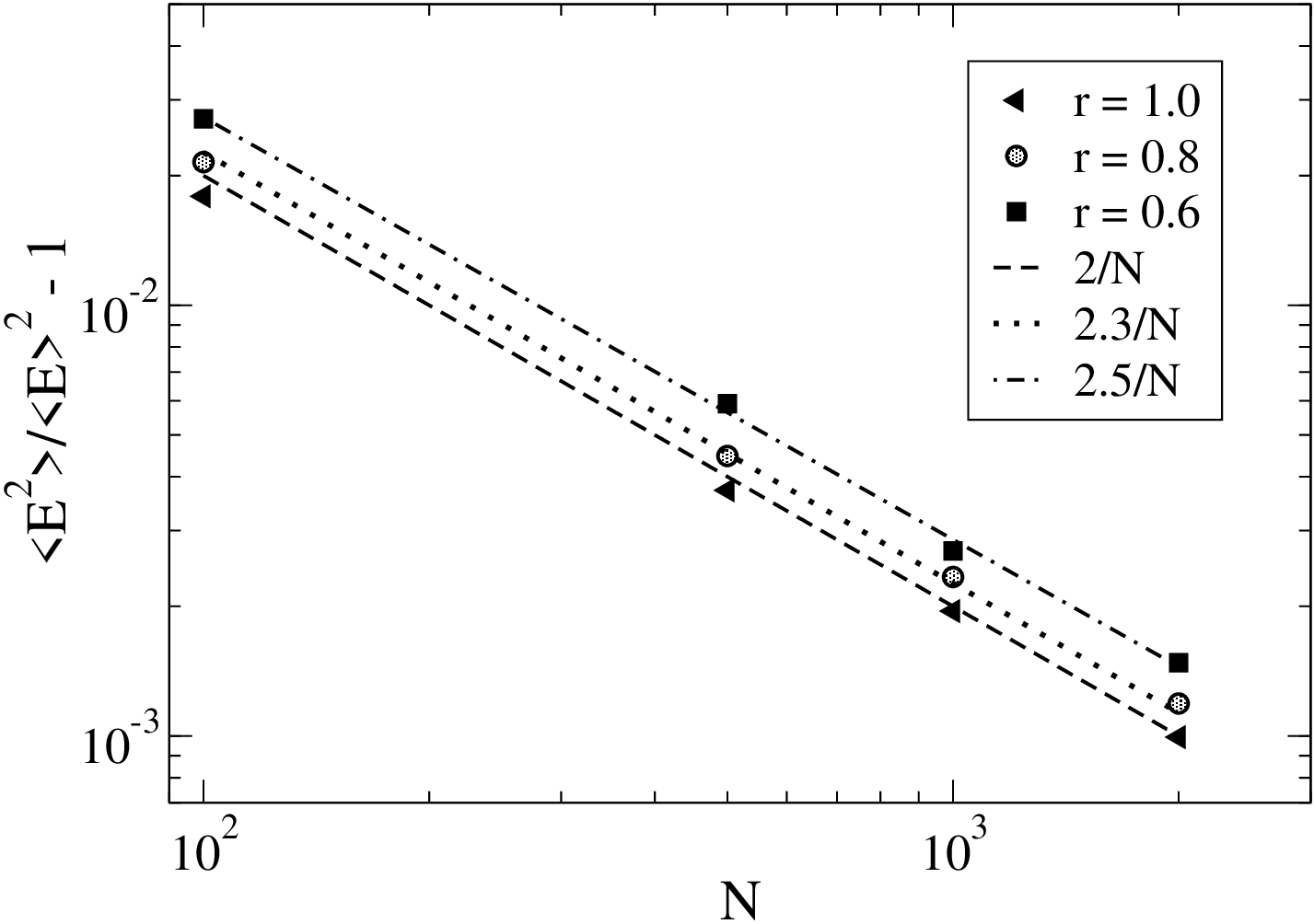}
\end{center}
\caption{Energy fluctuation as a function of the number
of particles, $N$ for $r=1.0,0.8,0.6$. 
The elastic case agrees with the theoretical prediction
$2/N$, whereas the inelastic case gives a value of the relative 
fluctuation slightly larger.
The remaining parameters are the same as in figure~\ref{velocitydistr}}
\label{tempflut}
\end{figure}

Interestingly, the energy fluctuations of our system, $E\equiv
m \sum_i v_i^2$/2, display a scaling with respect to the number
of particles. 
We are interested in the quantity
$({\langle E^2 \rangle -\langle E \rangle ^2)/\langle E \rangle^2}$ as a 
function of $N$ (at fixed density $\rho$). 
Defining $\langle v^n \rangle = \int v^n P(v) dv$, we have
\begin{equation}
\langle E^2 \rangle - \langle E \rangle^2 = 
\frac{m^2}{4} \sum_{i,j} \bigg( \langle v_i^2 v_j^2 \rangle
-\langle v_i^2 \rangle \langle v_i^2 \rangle\bigg)
\end{equation}
Under the hypothesis that
the variables $v_i$ are independently distributed, we get
\begin{equation}
\langle E^2 \rangle - \langle E \rangle^2 =
\frac{m^2}{4}{[N\langle v^4 \rangle  - N \langle v^2 \rangle^2]}
\end{equation}
Since for Gaussian variables $\langle v^4 \rangle = 3\langle v^2 \rangle^2$, 
we find
\begin{equation}
\frac{\langle E^2 \rangle - \langle E \rangle^2}{\langle E \rangle^2} = 
\frac{2}{N}
\label{enerflut}
\end{equation}
which is a well known formula for equilibrium systems
\cite{huang}. This scaling
is fairly well verified in figure~\ref{tempflut}. 

In the case of a granular fluid, $P(v_i)$ is no more Gaussian, exhibiting 
fatter tails, so one observes 
$\langle v^4 \rangle  \geq 3 \langle v^2 \rangle^2$ 
(for example in the case
$r=0.6$ we have $\langle v^4 \rangle / \langle v^2 \rangle \approx
3.3$). This leads to the conclusion that the scaling $\sim 1/N$
of formula~(\ref{enerflut}) still 
holds, but with a coefficient larger than $2$. 
Simulation runs for $r=0.6$ confirm this prediction (Fig.~\ref{tempflut}). 
The renormalization of the multiplicative constant occurring in the inelastic 
system could be
interpreted also as an ``effective reduction'' of the number of degrees of
freedom. Indeed, the inelastic system has the tendency to cluster, 
as it will be shown, and therefore the effective number of independent 
``particles'' appears smaller. Another
appealing interpretation is that the inelastic systems possesses an
effective ``specific heat'' larger than that of elastic systems.

\subsection{Velocity correlations}
\begin{figure}[htbp]
\begin{center}
\includegraphics[width=8cm,keepaspectratio, clip=true]{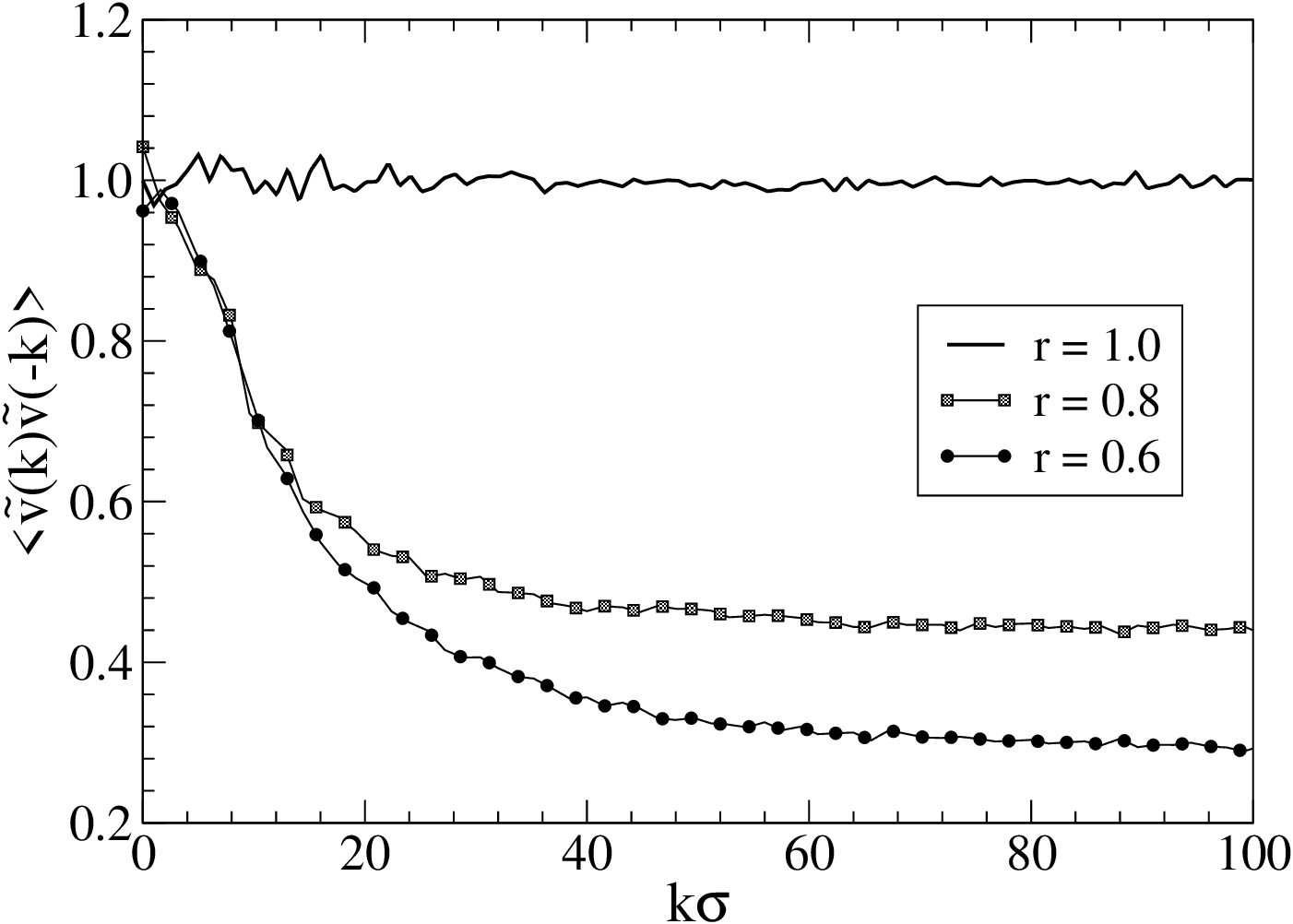}
\end{center}
\caption{
Structure function of the velocity field $v_i$, for elastic and
inelastic systems. The control parameters
are the same as in figure~\ref{velocitydistr}}
\label{velcor}
\end{figure}

A universal signature of the inelasticity is the
presence of correlations between the velocities of the particles.
We measured the structure function of the velocities $v_i$,
$S_v(k) =  \langle \tilde{v}(k)\tilde{v}(-k)\rangle$, where
$\tilde{v}(k)$ is the Fourier transform of $v_i$. 
In figure~\ref{velcor}, we show three $S_v(k)$ corresponding to the elastic 
($r=1$) and inelastic system with $r=0.8$ and $r=0.6$. 
As mentioned above,
the elastic systems is characterized by uncorrelated
velocities and this reflects on a constant structure function. 
A certain degree of correlation is instead evident in the inelastic
system. In fact, 
the inelasticity reduces by a
factor $r$ the relative velocity of two colliding particles 
and this leads to an increasing correlation among
velocities. However, the noise induced by the bath
competes with these correlations, making the structure function not very
steep. More specifically, $S_v(k)$ can be fitted, in the middle range of 
$k$ values, by an inverse power $\sim k^{-0.5}$, while at
high $k$ values it reaches a constant plateau. This is the fingerprint
of a persistent internal noise (velocity fluctuations are not
completely frozen by inelastic collisions).\cite{Ernstprogram}

\subsection{Distribution of interparticle spacing and of collision times}

\begin{figure}[htbp]
\begin{center}
\includegraphics[width=8cm,keepaspectratio, clip=true]{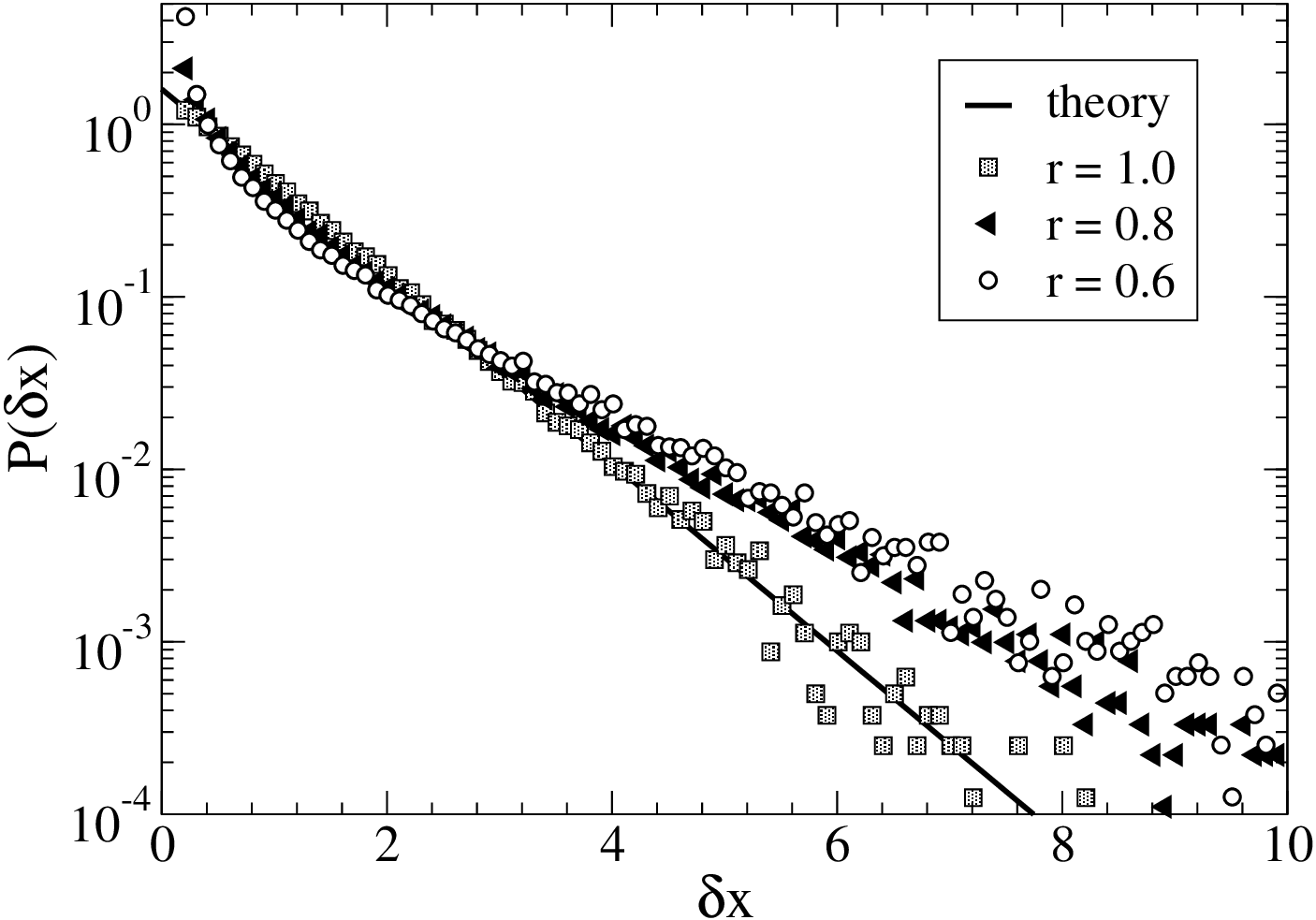}
\end{center}
\caption{
Distributions of distances between nearest
neighbor particles $\delta x = x_i - x_{i-1}$, for elastic
and inelastic systems, with the same parameters used before. The solid
line indicates the exponential expected in the elastic case (see
text). The state parameters are the same as in Fig.~\ref{velocitydistr}}
\label{spacingsdistr}
\end{figure}

The probability distribution, $P(\delta x)$, of distances
between nearest neighbor particles $\delta x = x_i - x_{i-1}$, shown in
figure~\ref{spacingsdistr}, provides
information about the spatial arrangement of
the system. In the elastic case one easily finds
\begin{equation} 
P(\delta x)=\frac{1}{\lambda} \exp[-(\delta x - \sigma)/\lambda]
\end{equation} 
for $\delta x \geq\sigma$ and $0$
for $\delta x <\sigma$ with $\lambda=(1-\rho\sigma)/\rho$. 
The presence of inelasticity modifies
such a simple exponential law 
in the way shown in figure~\ref{spacingsdistr}.  
In this case, the probability of finding two particles at
small separation increases together with that of finding
large voids. 
Such a picture is consistent with the idea of the
clustering phenomenon:\cite{zanetti} two particles, after the
inelastic collision, have a smaller relative velocity and therefore reach
smaller distances, eventually producing dense clusters and leaving 
larger empty regions with respect to the elastic case.

\begin{figure}[htbp]
\begin{center}
\includegraphics[width=8cm,keepaspectratio, clip=true]{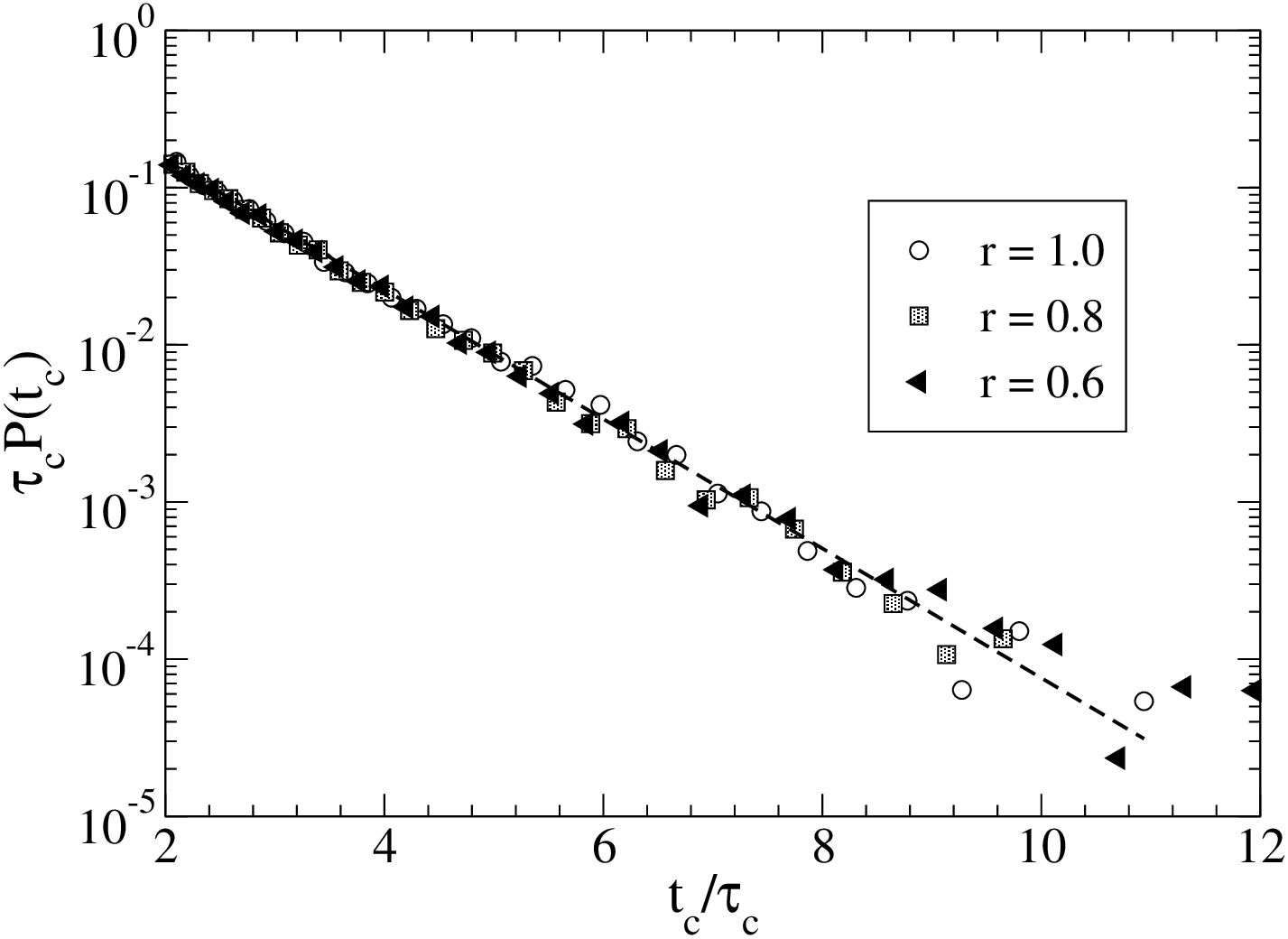}
\end{center}
\caption{Distributions of collision times. The dashed line indicates 
the exponential law expected for the elastic system (see text).
Control parameters as in figure~\ref{spacingsdistr}.}
\label{timesdistr}
\end{figure}

On the contrary, the probability distribution 
of collision times,
shown in figure~\ref{timesdistr}, appears to always follow the
theoretical (elastic) form $P(t)=1/\tau_c \exp(-t/\tau_c)$. Apart
from a trivial rescaling due to the change of the
thermal velocity with $r$, it seems not to depend appreciably
on the coefficient of restitution. 

Such a finding is in contrast with the situation observed
in 2D vibrated granular systems~\cite{Paolotti,Blair}. 
A possible explanation for this discrepancy is the following: 
in the inelastic 1D system, 
there is correlation between the relative velocities and the
free-paths (or free-times), otherwise the distribution
$P(t)$ and $P(\delta x)$ would have had the same shape 
due to the trivial relation $x=vt$, 
$v$ being the the average velocity of the rods. 
In particular, the fact that the peak
of $P(\delta x)$ in $\delta x = 0$ does not yield a corresponding 
peak in the $t=0$ region of $P(t)$ suggests that the shorter 
the distance between particles
the smaller their relative velocity.

\subsection{Density fluctuations}
We turn, now, to the study of
the structural properties of the inelastic hard-rod gas,
by considering
the pair correlation function or the structure factor.
As mentioned in section II, the virial equation~(\ref{pressure2})
relates the pressure of an elastic hard-rod system to its microscopic
structure.
In the presence of inelasticity, however, we expect 
that the tendency to cluster is mirrored by 
a change in the structural properties of the fluid.
Therefore we considered the behavior of the static (truly
speaking steady state) structure factor
\begin{equation}
S(k)=1+\rho\int dx [g(x)-1] e^{ikx}
\end{equation}
for different values of $\rho \sigma$ and inelasticity.

The spatial structure of the system is determined, as in ordinary
fluids, by the strong repulsive forces. Their role is seen in the oscillating
structure of $g(x)$. The inelastic nature of the collisions provides a 
correction to $g(x)$, which can be better appreciated by studying the small
wavelength behavior of $S(k)$ which develops a peak
at small $k$ recalling the Ornstein-Zernike behavior
\begin{equation}
S(k) \simeq \frac{1}{S^{-1}(0)+c_2 k^2}\;.
\end{equation}
The coefficient $c_2$ is negative for hard rods, whereas it is positive
for the inelastic system.
For hard rods, $S(k)$ is known~\cite{percus} and reads:
\begin{equation}
\label{sk_elastic}
S(k)=\frac{1}{1+2 b\rho\sigma \frac{\sin(k\sigma)}{k\sigma}
+( b\rho\sigma)^2 \frac{\sin^2(k\sigma/2)}{(k\sigma/2)^2}}
\end{equation}
with $b=\frac{1}{1-\rho \sigma}$.

\begin{figure}[htbp]
\begin{center}
\includegraphics[width=8cm,keepaspectratio, clip=true]
{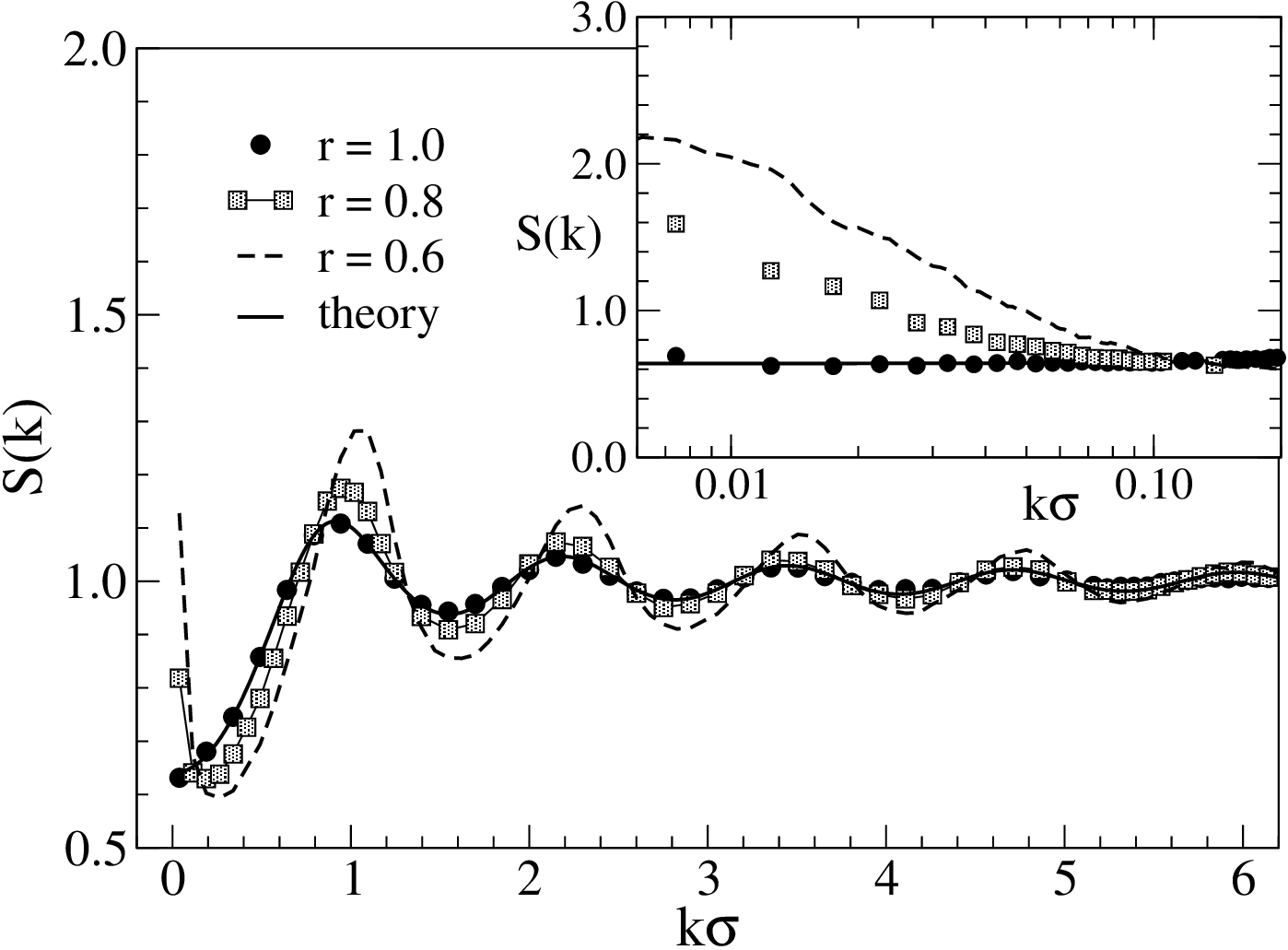}
\end{center}
\caption{Typical behavior of the structure function of $S(k)$
for different values of the  coefficients $r=1.0,0.8,0.6$. 
Notice the growth of the peak at small $k$ when $r$ decreases.}
\label{fig:sk}
\end{figure}
Figure~\ref{fig:sk} shows the typical behaviors of $S(k)$ for
elastic and inelastic systems. The numerically
computed structure factor of the elastic
system agrees rather well with  equation~(\ref{sk_elastic}). The 
inelastic system, instead, displays a peak in the small $k$ region 
reflecting the
tendency of the fluid to cluster. The peak increases 
with inelasticity, demonstrating that the energy dissipation in
collisions is responsible for these long range correlations. 
Incidentally, we comment that such a behavior of $S(k)$
could be attributed to the presence of a long range attractive effective
potential between the rods, as a result of dynamical 
correlations.\cite{ernst2} 

MackIntosh and Williams~\cite{mackintosh} found that, in the case
of randomly kicked rods in the absence of
viscosity, the pair correlation function decays as an inverse power law,
$g(x) \propto x^{-\eta}$ with $\eta
\to 0$ for $r \to 1$ and $\eta \to 1/2$ for $r \to
0$. Correspondingly, one expects $S(k)$ to diverge as $k^{-1/2}$ as
$k\to 0$, for very inelastic systems.  
In other words, the inelasticity
leads to long range spatial correlations which are revealed by the
peak at small $k$ of $S(k)$. We remark that, in spite of the apparent
similarity between the equations of state for elastic and inelastic system, 
their structural properties are radically different.  
Such a phenomenon is the result of the coupling of the long-wavelength
modes of the velocity field with the stochastic non-conserved driving force.
In fact, due to the inelastic collisions, the velocities of the particles 
tend to align, thus reducing the energy dissipation.
On the other hand, these modes adsorb energy from the heat bath
and grow in amplitude, and only the presence of friction prevents these 
excitations from becoming unstable. 
The density field, which is coupled to the velocity field by the continuity
equation, also develops long range correlations, and the structure
factor displays a peak at small wave-vectors.

\section{Transport properties}
One dimensional hard-core fluids exhibit an interesting connection
between the microscopic structural properties and diffusive ones.
In the present section, we present numerical results for the
collective diffusion and for the diffusion of a tagged particle
and show how these are connected to the structure.

\subsection{Collective Diffusion and Self-Diffusion}
Let us turn to analyze the perhaps simplest transport property of the
hard rods system, namely the self-diffusion, i.e. the dynamics of a
grain in the presence of $N-1$ partners. The problem is highly
non-trivial since the single grain degrees of freedom are coupled to
those of the remaining grains.  
Such a single-filing diffusion is also
relevant in the study of transport of particles in narrow 
pores.~\cite{kollmann}

The
diffusing particles can never pass each other. 
The excluded volume effect represents a severe hindrance for the particles 
to diffuse. In fact, a given particle in order to move must wait for a
collective rearrangement of the entire system. 
Only when the cage of a
particle expands, the tagged particle is free to diffuse
further.  This is a peculiar form of the so called cage effect 
which is enhanced by
the one-dimensional geometry.  In addition, the cage effect produces a
negative region and a slow tail in the velocity autocorrelation
function. 

As an appropriate measure of the self-diffusion,
we consider the average square displacement of
each particle from its position at a certain time, 
that we assume to
be $t=0$ without loss of generality
\begin{equation}
R(t)=\frac{\sum_{i=1}^N \langle [x_i(t)-x_i(0)]^2 \rangle}{N}
\label{eq:msd}
\end{equation}

At an early stage, the self-diffusion is expected to
display ballistic behavior, $R(t) \sim |v|^2t^2$, with
$|v|^2=T_g/m$, before any perturbation (heat bath and collisions) change
the free motion of particles, i.e. when $t \ll 1/\gamma$ and $t \ll
\tau_c$. 

\begin{figure}[htbp]
\begin{center}
\includegraphics[width=8cm,keepaspectratio, clip=true]
{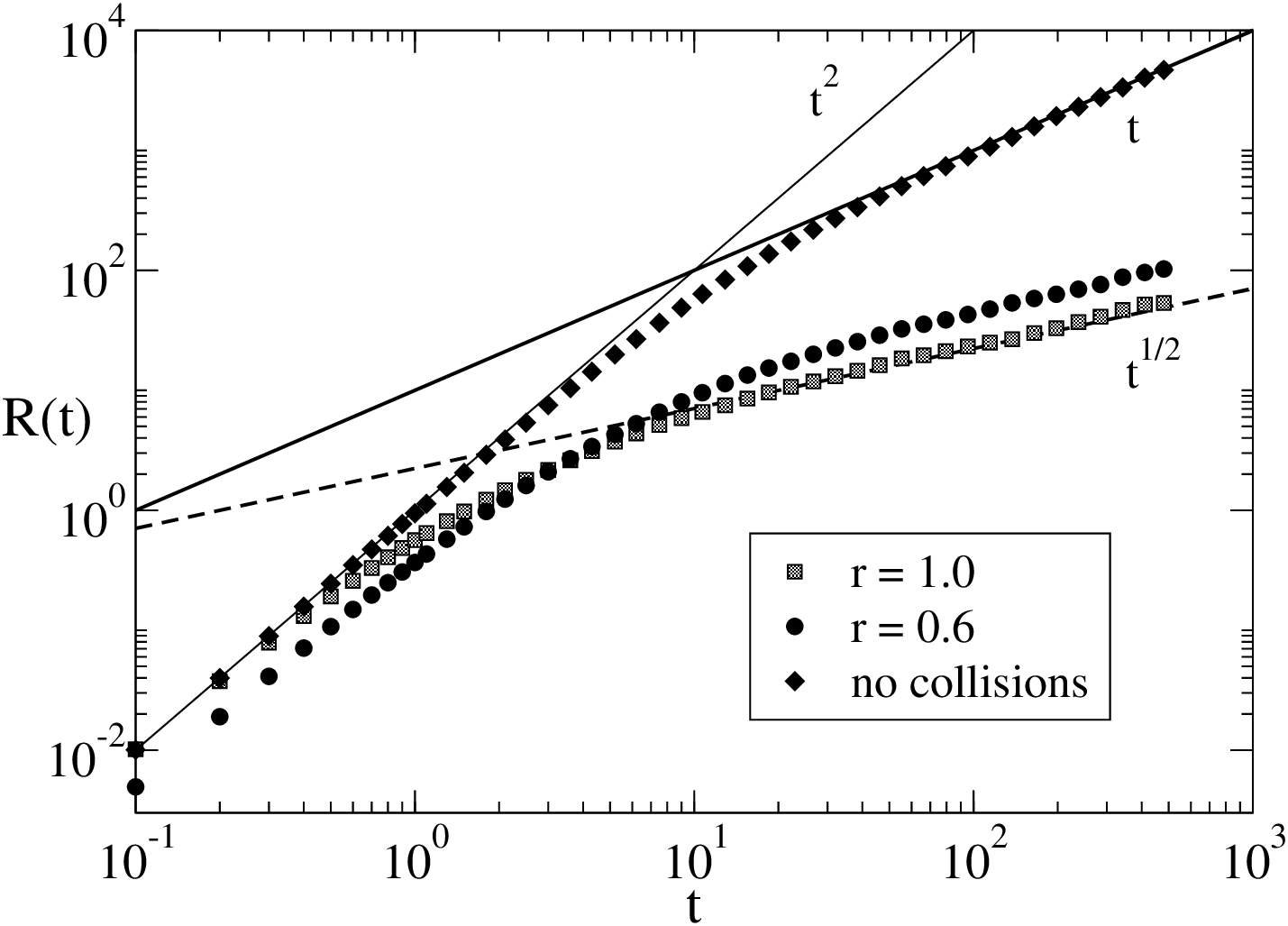}
\end{center}
\caption{Self-diffusion: behavior of $R(t)$ for
three different systems: one without collisions 
(free particles), one with elastic
collisions and the third with inelastic collisions ($r=0.6$).}
\label{fig:diff}
\end{figure}

A system of non-interacting (i.e. non-colliding) particles subjected
to Kramers' dynamics~(\ref{kramers1},\ref{kramers2}) displays, after 
the ballistic
transient, normal self-diffusion of the form $R(t) \sim 2D_0t$ with
$D_0=T_g/\gamma$. This is well verified in figure~\ref{fig:diff}
(circles).

Lebowitz and Percus~\cite{percuslebow} studied the tagged particle
diffusion problem for systems governed by non-dissipative dynamics
without heat bath and found a diffusive behavior described by
\begin{equation} 
R(t)=2 D_{coll} t= \lambda \langle |v| \rangle t
\label{eq:percus}
\end{equation}
with $\lambda=(1-\rho\sigma)/\rho$.
Since $\langle |v| \rangle = \sqrt{T/2 \pi m}$ one obtains
\begin{equation} 
\label{eq:fabia}
D_{coll}=\frac{1-\rho \sigma}{\rho}\sqrt{\frac{T}{2\pi m}}
\end{equation}

On the other hand, almost in the same years, Harris~\cite{harris}
studied the behavior of $R(t)$ in the case of $N$ identical Brownian
particles with hard-core interactions , i.e. obeying a single-filing
condition, and obtained a sub-diffusive behavior increasing as
\begin{equation}
R(t)=2 \lambda \bigg(\frac{D_0 t}{\pi}\bigg)^{1/2}
\label{singlefile}
\end{equation}
where $D_0$ is the single (noninteracting) particle diffusion coefficient. 

In figure~\ref{fig:diff}, we study the self-diffusion $R(t)$ for
elastic and inelastic systems in the presence of heat bath and
viscosity, obtaining two different regimes separated by a
typical time $\tau_c$. In the first transient regime $t<\tau_c$ we
we observe the ballistic motion.
In the second stage, instead, we
expect the sub-diffusive behavior, $R(t)\sim t^{1/2}$, 
predicted by Harris and
other authors.~\cite{harris,pincus,olandesi}. The inelastic
system displays the same sub-diffusive behavior, but with a
multiplicative constant larger than $1$, i.e. at equal times the
granular (inelastic) fluid has a larger absolute value of $R(t)$.

It is interesting to analyze the connection between this transport
property and the compressibility of the system, as remarked by
Kollmann.~\cite{kollmann}

\subsection{Connection between self-diffusion and structure}
We follow Alexander and Pincus~\cite{pincus} argument
in order to derive a formula for the self-diffusion and show the
connection with the compressibility of the system.  Let us consider
the two time correlator:
\begin{equation}
\langle \rho_k(t)\rho_{-k}(0) \rangle =
\sum_{ij} \langle e^{i k x_i(t)}e^{-i k x_j(0)} \rangle
\end{equation}
we set $x_i(t) = X_i+(x_i(t)-X_i) = X_i+u_i$, where $X_i$ are the nodes
of the 1D lattice $X_i=i a$, ($a$ being the lattice spacing). 
Expanding around $X_i$ 
\begin{equation}
\langle \rho_k(t)\rho_{-k}(0) \rangle \simeq k^2 
\sum_{ij}\mbox{e}^{ik(X_i-X_j)}\langle u_i(0) u_j(t) \rangle 
\end{equation}
hence
\begin{equation}
\langle \hat u_k(t) \hat u_{-k}(0) \rangle = 
\frac{\langle\rho_k(t)\rho_{-k}(0)\rangle}{N k^2}
\label{aggiunta}
\end{equation}
We assume, now, that the density correlator varies as 
\begin{equation}
\langle \rho_k(t)\rho_{-k}(0)\rangle = 
\langle \rho_k(0)\rho_{-k}(0)\rangle e^{-Dk^2 t}
\end{equation}
where $D=D_0/S(0)$ (this is demonstrated in Appendix A) is the
collective diffusion coefficient, with $D_0=T_b/\gamma$.  
Now, the mean square displacement per particle~(\ref{eq:msd})
can be written as
$
R(t) = \sum_l \langle [u_l(t)-u_l(0)]^2 \rangle/N  
$
and, in Fourier components, reads
\begin{equation}
R(t) = \frac{2}{N} \sum_k\langle \hat u_k(0) \hat u_{-k}(0) - 
\hat u_k(t) \hat u_{-k}(0) \rangle 
\end{equation}
employing Eq.~\ref{aggiunta} we find
\begin{equation}
R(t) = 
2 \sum_k \langle \rho_k(0)\rho_{-k}(0) \rangle 
\frac{1-{\mbox e}^{-Dk^2 t}}{N^2 k^2}
\end{equation}
Approximating the sum with an integral 
($\sum_k  \to L/(2\pi) \int_{-\pi/a}^{\pi/a} dk$
and recalling that 
\begin{equation}
N S(k)= \langle \rho_k(0)\rho_{-k}(0)\rangle 
\end{equation}
we obtain
\begin{equation}
R(t) = 
\frac{2L}{N}\int_{0}^{\pi/a} \frac{dk}{2\pi} 
S(k) \frac{1-\mbox{e}^{-Dk^2 t}}{k^2}
\end{equation}
and therefore
\begin{equation}
R(t) \simeq
\frac{2}{\rho \pi} S(0) \sqrt{\pi D t}
=\frac{2}{\rho} \sqrt{\frac{D_0 S(0) t}{\pi}}
\end{equation}
Notice that such a formula in the case of hard-rods is identical
to formula~(\ref{singlefile}).

We see that the tagged particle diffusion depends on the structure of
the fluid. In the granular fluid the $k \to 0$ part of the spectrum is
enhanced and thus we expect a stronger tagged particle diffusion. This
is what we observe. Physically there are larger voids and particles can
move more freely. Let us notice that as far as the collective
diffusion is involved the spread of a group of particles is faster in
the presence of repulsive interactions than
without.~\cite{tarazona}

\section{Conclusions}
In this paper 
we have studied a one-dimensional system of inelastic hard-rods coupled
to a stochastic heat-bath with the idea
that it can represent a reference system
in the area of granular gases to test theories
and approximations. Due to the relative simplicity
of the one dimensional geometry we have shown that it is possible
to obtain relations between the macroscopic control parameters 
such as kinetic temperature, pressure and density. 
We tested these analytical predictions against the numerical measurements and found a fairly good
agreement.
It also appears that many properties of the 
heated one-dimensional inelastic hard rod system
are similar to those of ordinary fluids.
However, when we have considered how various physical observables 
fluctuate about their equilibrium values, many relevant differences
have emerged. These range from the non-Gaussian behavior of the 
velocity distribution, the peculiar 
form of the distribution of distances between particles and of  
the energy fluctuations to the shape of the structure factor
at small wave-vectors. Finally, we have found that the
diffusive properties of the system are affected by the
inelasticity and in particular the self-diffusion is 
enhanced.

To conclude, in spite of the similarity between ordinary
fluids and granular fluids, which has been
recognized for many years and has made possible
to formulate hydrodynamical equations for
granular media in rapid, dilute flow, the presence
of anomalous fluctuations in the  
inelastic case indicates the necessity of 
a treatment which incorporates in a proper way 
both the local effects
such as the excluded volume constraint and
the long ranged velocity and density correlations.
Such a program has been partially carried out by Ernst and coworkers
\cite{Ernstprogram}, but needs to be completed regarding the
description of the fluid structure.

\section{Appendix A}
In the case of over-damped dynamics (i.e. large values of $\gamma$)
one finds that the collective
diffusion is given by~\cite{tarazona}
\begin{equation}
\frac{\partial \rho(x,t)}{\partial t} = \frac{1}{\gamma}
\frac{\partial}{\partial x} \bigg\{\rho(x,t)\frac{\partial 
\mu(\rho(x))}{\partial x}\bigg\}
\label{collective}
\end{equation}
where $\mu(\rho(x))$ is the local chemical potential.
Expanding $\mu$ about its average value $\rho_0$ we obtain:
\begin{equation}
\frac{\partial \mu(\rho(x))}{\partial x}
=\bigg[\frac{\delta \mu}{\delta \rho}\bigg]_{\rho_0}
\frac{\partial \rho(x)}{\partial x}
\end{equation}
Substituting into Eq.~(\ref{collective}) we find
\begin{equation}
\frac{\partial \rho(x,t)}{\partial t}=\frac{1}{\gamma}
\rho_0\bigg[\frac{\delta \mu}{\delta \rho}\bigg]_{\rho_0}
\frac{\partial^2 \rho(x)}{\partial x^2}
\end{equation}
and using
\begin{equation}
S(0)=\frac{K_B T_b}{\rho_0} 
\bigg[\frac{\partial\rho_0}{\partial\mu}\bigg]_T
\end{equation}
in the case of elastic hard rods we obtain:

\begin{equation}
\frac{\partial \rho(x,t)}{\partial t}=\frac{1}{\gamma}
\frac{K_B T_b}{S(0)}
\frac{\partial^2 \rho(x)}{\partial x^2}
=\frac{D_0}{S(0)}\frac{\partial^2 \rho(x)}{\partial x^2}
\end{equation}

Thus the renormalized diffusion coefficient is $D=D_0/S(0)$.



\end{document}